\documentclass{article}
\pdfoutput=1 

\usepackage[preprint]{neurips_2024}

\usepackage[utf8]{inputenc} 
\usepackage[T1]{fontenc}    
\usepackage{hyperref}       
\usepackage{url}            
\usepackage{booktabs}       
\usepackage{amsfonts}       
\usepackage{nicefrac}       
\usepackage{microtype}      
\usepackage{xcolor}         
\usepackage{graphicx}
\usepackage{listings}
\usepackage{longtable}
\usepackage{tabularx}

\newcommand*{\affaddr}[1]{#1}
\newcommand*{\affmark}[1][*]{\hspace{0.085em}\textsuperscript{#1}}

\title{What do MLLMs hear? Examining reasoning with text and sound components in Multimodal Large Language Models}

\author{%
  Enis Berk \c{C}oban\affmark[1] \\
  \texttt{ecoban@gradcenter.cuny.edu}
  \And
  Michael I Mandel\affmark[1] \\
  \texttt{mim@mr-pc.org}
  \And
  Johanna Devaney\affmark[1,2] \\
  \texttt{jdevaney@gc.cuny.edu}
 \AND
  \vspace{-5mm}\\
  \affaddr{\affmark[1] The Graduate Center, CUNY }\hspace{12pt}
  \affaddr{\affmark[2] Brooklyn College, CUNY}\hspace{12pt}
}

\hypersetup{
pdftitle={What do MLLMs hear? Examining reasoning with text and sound components in Multimodal Large Language Models},
pdfauthor={Enis Berk \c{C}oban, Michael I Mandel, Johanna Devaney},
pdfkeywords={MLLMs, reasoning, Multimodal Large language models}
}

\begin{document}

\maketitle

\begin{abstract}
Large Language Models (LLMs) have demonstrated remarkable reasoning capabilities, notably in connecting ideas and adhering to logical rules to solve problems. These models have evolved to accommodate various data modalities, including sound and images, known as multimodal LLMs (MLLMs), which are capable of describing images or sound recordings. Previous work has demonstrated that when the LLM component in MLLMs is frozen, the audio or visual encoder serves to caption the sound or image input facilitating text-based reasoning with the LLM component. We are interested in using the LLM’s reasoning capabilities in order to facilitate classification. In this paper, we demonstrate through a captioning/classification experiment that an audio MLLM cannot fully leverage its LLM’s text-based reasoning when generating audio captions. We also consider how this may be due to MLLMs separately representing auditory and textual information such that it severs the reasoning pathway from the LLM to the audio encoder. 
\end{abstract}

\section{Introduction}
\label{intro}

Humans can learn from descriptions of events or ideas and can recognize them afterwards, even if they are observing such an event for the first time. Can the reasoning abilities in large language models (LLMs) enable them to achieve a similar goal? It has recently been shown that LLMs trained on internet-scale data have zero and few-shot capabilities \citep{brown2020language,kojima2022large}, demonstrating that they can solve tasks for which they were not specifically trained. For example, LLMs trained to predict the next chunks of text in a given text can also perform other natural language tasks, such as summarizing a given text. More complex tasks that LLMs cannot solve from scratch can be solved by in-context learning, where question and answer pairs are provided in the prompt, which act like training data. Another way to leverage in-context learning is by providing related information in the prompt to help the model generate solutions based on the given information. LLMs can identify and connect related ideas in the given text, and draw conclusions regarding these facts or ideas. Reasoning abilities allow LLMs to make connections between related concepts and provide responses by collating different information present in their training data \citep{wei2022emergent}, although there is a debate around whether these abilities are emergent in larger-scale models \citep{schaeffer2023emergent}.  While reasoning abilities are present in LLMs, they are limited, both due to catastrophic forgetting, which causes reasoning to disappear \citep{de2021continual}, and hallucinations, which leads to fallacy generation \citep{tonmoy2024comprehensive}.

Multimodal large language models (MLLMs), where the output of image or audio encoders are tokenized and input into an LLM, exhibit some of the reasoning capabilities found in non-multi-modal LLMs \citep{wang2024exploring}. We are interested in leveraging their reasoning ability to classify low-resource classes using only descriptions of the given classes for both zero- and few-shot learning. For the latter, MLLMs could use text-based reasoning to not only learn from a small number of labeled samples but also generalize better, such that the presence of unrelated data in samples, like background noise, would not impact classification. Previous demonstrations of MLLM's reasoning abilities have been reasoning \emph{about} the image or audio input \citep[e.g.,][]{gong2023listen}. 
In order to leverage the reasoning to learn to classify unseen audio from textual descriptions, the MLLM needs to co-reason \emph{with} the multimodal content. Research on vision MLLMs, however, has demonstrated that they do not possess these co-reasoning capabilities, that the models depend on the input modalities to control output as if they are simple flags turning on and off a specific task \citep{qi2023limitation}. In this paper, we analyze audio MLLMs capabilities in order to more fully understand their co-reasoning abilities and limitations, and to discuss possible solutions to this problem.

\section{Reasoning in multimodal large language models}
\label{reasoningMLLM}
The earliest MLLMs had encoders that output embeddings representing each input modality, which were then combined using different strategies, such as simple combinations, before being fed into another language model \citep{wu2023multimodal}. Initial models used different encoders for each modality, depending on what was most commonly used for such input (e.g., CNNs for images and RNNs for text \citep{you2016image}). The subsequent development and success of transformers on different modalities led to a uniform encoder approach, with similar attention architectures and sizes, that is commonly used in current MLLMs \citep{radford2021learning}. This has facilitated the development of MLLMs that can incorporate images and/or audio with the LLM core text-based functionality. 

\begin{figure}
  \centering
  \includegraphics[width=0.8\textwidth]{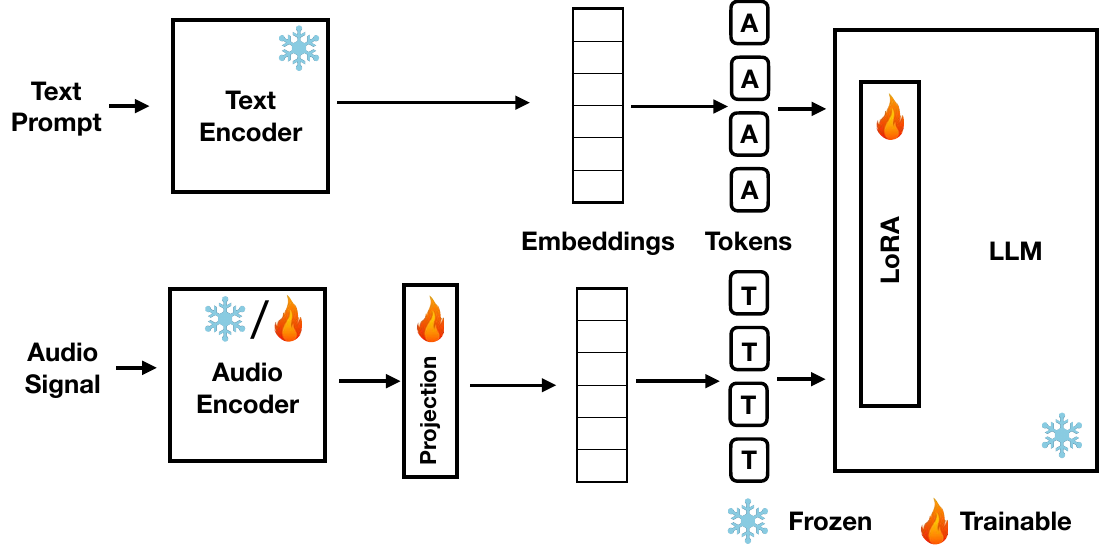}
  \caption{Generic audio MLLM architecture, specific components may vary with specific models. The snowflake represents components that are typically frozen and the flame represents those that are typically trained (or can be in the case of the `/').}
  \label{audioLLMfigure}
\end{figure}

\subsection{Visual reasoning in MLLMs}
\label{visionMLLMs}
Much of the work on reasoning in vision MLLMs builds on the Visual Question Answering (VQA) \citep{Antol_2015_ICCV}, a captioning task that arguably requires a model to demonstrate complex reasoning capabilities. Modifications to the original VQA approach include using different metrics that better reflect real-world visual concepts \citep{Kervadec_2021_CVPR} and visualization approaches that allow for finer-grained investigation of vision MLLMs' reasoning capabilities \citep{jaunet2021visqa}. Recent evaluations on visual reasoning have shown that the representation of visual information in MLLMs is a bag of words, rather than an ordered representation, as evidenced by the inability of the models to answer any questions related to the order of the objects in the images \citep{yuksekgonul2022}. Vision MLLMs also lack spatial reasoning capabilities when queried about the left-right location of objects in an image \citep{kamath2023s}. Another problem with vision MLLMs is their lack of understanding of relationships between objects, such as when they incorrectly assign human actions to animals, and vice-versa \citep{thrush2022winoground}.  Research into visual reasoning capabilities in vision MLLMs is facilitated by the development of benchmarks, which have demonstrated some of their ongoing reasoning and hallucination issues \citep{fu2023mme, fan2024nphardeval4v}. These include vision MLLMs performing worse than LLMs on instruction following \citep{zeng2023matters} and vision MLLMs exhibiting a lack of semantic grounding, which limits their ability to leverage textual relationships present in the LLM in the visual modality \citep{lu2024evaluation}.

\citep{kervadec2019weak,song2023bridge} to facilitate reasoning on more fine-grained aspects of the data. While generating more samples is a sufficient solution for specific abilities, it requires having matching data samples from each modality. As a result, it becomes even harder to come up with sufficient training data for data-hungry deep-learning models. A well-aligned model would be able to use abilities gained in one modality in another without requiring any extra data. For example, if a model can converse about modes of transportation, it should be able to show similar reasoning capabilities given an image of a car. 

In-context learning through prompts is used to facilitate and analyze reasoning in MLLMs \citep{zhao2023mmicl}. Reasoning abilities typically improve when a model is forced to take specific steps through its prompt \citep{kojima2022large, yao2022react} and prompt probing (including prompts related to visual, textual, and outside information) has been key in the understanding of MLLMs visual reasoning limitations \citep{qi2023limitation}. Such probing has shown that in MLLMs, the use of non-linguistic prompts can increase the risk of catastrophic forgetting \citep{wang2023paxion}, reducing any reasoning capabilities that they do exhibit. It also underlines the importance of incorporating outside knowledge in the testing paradigm when assessing reasoning \citep{marino2019ok} as a way to assess the contribution of how much text-based reasoning concepts in the language model are being leveraged in the MLLM. Typically multi-step training is employed to mitigate catastrophic forgetting, the second modality is aligned to the frozen LLM and then fine-tuned with LLM low-rank adaption (LoRA) \citep{alayrac2022flamingo, ye2023mplug}. 

\subsection{Audio reasoning in MLLMs}
\label{audioMLLMs}

Audio MLLMs are a more recent development than vision MLLMs \citep[][e.g.,]{deshmukh_pengi_2023, silva2023collat, gong2023listen, tang_salmonn_2023, tang2024avicuna}.  Arguably, the first audio MLLM was Pengi \citep{deshmukh_pengi_2023}, which frames all audio tasks as a text generation task in order to leverage the causal capabilities of the integrated LLM on the input audio prompt. It uses the hierarchical audio transformer HTS-AT \citep{chen2022hts} for its audio encoder, CLIP \citep{radford2021learning} for its text encoder, and applies contrastive pre-training with CLAP \citep{elizalde2023clap}. While \citet{deshmukh_pengi_2023} evaluate Pengi on a range of open-ended tasks, such as audio captioning and audio question/answering, and closed tasks related to classification and retrieval, they do not evaluate Pengi's reasoning capabilities. Some slightly more recent audio MLLMs that explore the concept of reasoning are Listen, Think, Understand (LTU) \citep{gong2023listen} and SALMONN \citep{tang_salmonn_2023}. Figure \ref{audioLLMfigure} provides a generalized overview of the architecture of these models.

LTU uses Audio Spectrogram Transformer (AST) \citep{gong2021ast} for its audio encoder and the LLaMA text encoder \citep{touvron2023llama} along with CAV-MAE \citep{gong2022contrastive} for contrastive pre-training. Like Pengi, LTU treats all audio as input for automatic audio captioning (AAC), from which it leverages the reasoning abilities of LLaMA to reason \emph{about} the audio captions. They freeze AST and use Low-rank Adaptation (LoRA) \citep{hu2021lora} to force the model to condition on the audio captions and not rely simply on the language model, which helps to minimize hallucinations. In evaluating LTU's reasoning capabilities, \citet{gong2023listen} are concerned with the model's ability to ``think'', which the authors argue is demonstrated by tasks where the model explains an audio caption, and ``understand'', which they further argue is demonstrated by tasks where the model has to infer further action. \citet{gong2023joint} also released LTU-AS, which integrates a speech model.

SALMONN uses two audio encoders one from a speech model and one from a generic audio model, which gives it automatic speech recognition (ASR) capabilities, like LTU-AS. SALMONN feeds the output of Whisper's speech encoder \citep{radford2023robust} and BEAT's audio encoder \citep{chen2023beats} for generic audio sound into Q-Former query transformer \citep{li2023blip} to generate audio tokens for input into Vicuna \citep{chiang2023vicuna}.  \citet{tang_salmonn_2023} specifically evaluate the effect of fine-tuning on the reasoning tasks. Since most of their data consists of ASR and AAC instructions, the model tends to ignore the prompt and respond with one of these answers. To address this, activation tuning is applied, which is lowering the scaling factor of the LoRA method. As with Pengi and LTU, the reasoning tasks in SALMONN are performed on the text generated from the audio samples, either captions or transcribed speech. Evaluation of SALMONN revealed a similar phenomenon to those observed in visual MLLMs, where the model forgets some of the text-based commonsense knowledge available in the LLM.

Evaluations of reasoning in audio MLLMs, primarily with LTU and SALMONN, have examined and proposed ways to address the observation made in visual MLLM literature that MLLMs exhibit issues with instruction following compared to LLMs. The lack of semantic grounding, however, has not been evaluated and is the focus of this paper. 

\section{Experiment 1: In-context audio classification}
\label{exp1}

Audio MLLMs have demonstrated competence, if not state-of-the-art performance, on in-context classification tasks. These models, when provided with a minimal set of examples or a succinct description of the anticipated output, are capable of discerning the underlying structure and accurately completing the given input. In multimodal systems, in-context learning extends to interactions with image or audio inputs, broadening the model's understanding to other modalities. This enables the model to comprehend examples that reference the subject of interest in an image or audio context. As a result, even when provided solely with visual or auditory descriptions of expected labels in-context, the model can effectively undertake classification tasks across different modalities, suggesting a form of reasoning. This capability proves particularly advantageous for resource-constrained classes, such as rare bird species that have detailed written reports on their calls and songs \citep[e.g.,][]{hannon1998willow}. Audio MLLMs, leveraging their in-context learning proficiency, could effectively harness this written data to enhance classification performance in low-resource scenarios.

\subsection{Methodology}
\label{exp1methodology}

We used the pre-trained LTU model as our base model in these experiments. The LTU model was trained using existing audio datasets, their labels, descriptions, transcripts, and metadata as available. As described in Section \ref{audioMLLMs}, LTU generates text captions for an input sound. These captions can be utilized for classification by calculating the similarity between them and the expected labels. This similarity is then interpreted as a confidence score for each label. We adopted the methodology outlined in the LTU paper for this task, employing OpenAI's text-embedding-ada-002 model to obtain text embeddings for both the labels and model outputs. For each sample, we computed the cosine similarity between the label embeddings and the model's output or caption. These similarity scores were then used as confidence indicators for the corresponding label of the given data point.

The advantage of relying on the similarity between text embeddings is that it enables the matching of related captions with the correct label, even if the given label does not appear in the caption. However, a potential disadvantage is the loss of semantic information, which could lead to incorrect label matching due to the diverse information contained within the caption. In order to quantify this, we conducted the same experiments repeatedly, with the variance being 0.7\%, a figure negligible in comparison to the changes in results---up to 20\%---observed between different experiments, as demonstrated in Tables \ref{incontext} and \ref{ltuClassification}.

We ran this experiment on EDANSA \citep{ccoban2022edansa}, a bioacoustics audio dataset collected in Alaska, which comprises 10,782 10-second samples, totaling 27 hours of audio. This multi-label dataset contains 28 distinct labels, organized in a hierarchical structure. We selected 12 labels that represent main events and have more than 400 samples in total, namely: Anthrophony, Aircraft, Biophony, Songbird, Bird, Grouse, Insect, Waterfowl, Geophony, Rainfall, Wind, and Silence. Since the LTU model does not use EDANSA, nor any other ecological soundscapes, in its training set, it is an ideal choice for testing LTU's out-of-distribution performance. Furthermore, labels such as Geophony and Biophony are not common in LTU's training set, allowing for better evaluation of the effect of fine-tuning.

While LTU has been shown to underperform compared to supervised approaches for classification tasks \citep{gong2023listen}, we are interested in its in-context classification capabilities using knowledge from the text domain. Therefore, we tested the original LTU model on EDANSA labels to establish a baseline performance. In this experiment, we employed the most effective prompt from the original LTU classification experiments: `write an audio caption describing the sound.' We noted that the model often failed to output expected labels such as biophony due to a lack of examples in the training. Consequently, we devised alternative labels (Appendix \ref{alternative_labels}) that are synonymous and incorporated these into our experiments.

However, even if using similarity measures aids in classification, LTU's open-ended captions may not always match expected labels, even after fine-tuning. This is a limitation compared to supervised models with fixed output labels, as the model might not use exact labels or might miss silent or secondary sound events. To address this, we included expected labels in the prompts, such as ``Write an audio caption describing the sound. Could the sound be \{list of comma separated labels\}?".

Subsequently, we fine-tuned the LTU model on the EDANSA dataset using the suggested train, test, and validation split from the original paper. We employed LoRA to fine-tune the language model, and for the audio encoder, we experimented with full fine-tuning and only training the projection layer. With LoRA, we used a learning rate of 16, as suggested by the original LTU paper \citep{gong2023listen}, and also experimented with values of 2, 4, and 8. We opted for lower values based on research indicating that a lower learning rate enhances performance by preventing catastrophic forgetting \citep{tang_salmonn_2023}. We used a batch size of 256, a LoRA alpha of 1, and a LoRA dropout of 0.05. We applied early stopping if validation did not improve after 10 epochs and used the checkpoint with the lowest validation score.

For these experiments, we primarily used the 13B version of the LTU model. The in-context learning experiments in this Section~\ref{exp1} were run with both 7B and 13B models, but we found the performance of the 13B model to be superior, and thus its performance is reported for all experiments.

To conduct these experiments, we utilized two NVIDIA A40 GPUs, each with 48GB of memory. Each fine-tuning run took approximately 3 hours, and we conducted a total of 46 runs. This amounted to 11.5 days of GPU time in total.

\subsubsection{In-Context Learning with Grouse Call Descriptions}

Of all the labels we examined, Grouse proved to be the most formidable challenge in terms of classification. This is primarily due to the limited number of samples available for this label, coupled with the complex and diverse array of calls that the Grouse is known to produce. The grouse featured in the EDANSA dataset, several species of ptarmigan, is renowned for its wide range of sounds, including hissing, chirping, and peeping. However, it is most notably recognized for the drumming sounds produced by the male \citep{hannon1998willow}. As a result of this complexity and diversity, supervised models encounter significant difficulty when classifying the Grouse label. These models typically require a larger sample size when dealing with a class that exhibits a high degree of diversity within its samples. Given the known descriptions of these specific calls and the limited number of samples, we hypothesized that the LTU model would effectively leverage the written information and outperform the supervised model.

To test this hypothesis, we incorporated descriptions of the grouse's calls into the prompt, asking the model to identify these sounds within the given audio. It's important to note that we only provided this additional information during the test and validation phases. The prompt for this experiment reads as follows: ``Provide labels for the audio file. Could it be a Grouse? Here is detailed information on Grouse call types: \ldots (3) Krrow is a medium-length (50-300 ms) call that rises quickly and falls slowly in frequency: In males, it sounds like `bugow'; in females, like `meow'; it is typically given during aggressive disputes \ldots Please list the labels for the audio file." A comprehensive version of all call types is provided in the supplemental materials in Appendix \ref{SuppPrompt}. 

\subsection{Results}
\label{exp1results}
\begin{table}
  \caption{LTU classification performance versus a supervised model. `Vanilla' is un-fine-tuned LTU, `Partially Fine-tuned' is with tuning the audio projection, `Fully Fine-tuned' is tuning the audio encoder and applying LoRA training. Mean AUC and F1 are computed across all EDANSA classes.}
  \label{ltuClassification}
  \centering
  \begin{tabular}{lll}
    \toprule
    Experiment     & Mean AUC & Mean F1 \\
    \midrule
    Supervised & 0.97 & 0.90 \\
    \midrule
    Vanilla &  0.69 & 0.54  \\    
    Partially Fine-tuned & 0.72 & 0.59 \\
    Fully Fine-tuned & 0.81 & 0.68 \\
    \bottomrule
  \end{tabular}
\end{table}

\begin{table}
  \caption{In-context prompting results, compared to a supervised model and fully fine-tuning (fine-tuning the audio encoder and applying LoRA training to the LLM). `Label in Prompt' refers to explicitly listing all class labels in the prompt and `Grouse Call in Prompt' refers to providing acoustic information about the Grouse call in the prompt.} 
  \label{incontext}
  \centering
  \begin{tabular}{lcc@{$\qquad$}cc@{$\qquad$}cc}
    \toprule
    Experiment & \multicolumn{2}{c}{Mean$\qquad$}  & \multicolumn{2}{c}{Grouse$\qquad$}  & \multicolumn{2}{c}{Insect} \\
     & AUC & F1 & AUC & F1 & AUC & F1\\    
    \midrule
    Supervised & 0.97 & 0.90 & 0.97 & 0.80 & 0.97 &  0.96\\
    Fully Fine-Tuned & 0.81 & 0.68 & 0.82 & 0.41 & 0.94 & 0.95\\
    \midrule
    Labels in Prompt & 0.81 & 0.68 & 0.79 & 0.35 & 0.93 & 0.81\\
    Grouse Call in Prompt & 0.81 & 0.68 & 0.76 & 0.46 & 0.93 & 0.82\\

    \bottomrule
  \end{tabular}
\end{table}

The results of our audio classification experiments are encapsulated in Table \ref{ltuClassification}. As anticipated, the LTU model's performance in classifying samples from the EDANSA dataset falls short when compared to the supervised model. The supervised approach achieves a mean AUC score across labels of 0.97, whereas the LTU model lags considerably behind with a mean AUC of 0.69. Upon training the audio projection layer (``Partially Fine-tuned"), which serves as the intermediary between the LLM and the audio encoder, the performance experiences a marginal increase of 0.03 points. This layer is instrumental in aligning audio embeddings with input tokens, and the modest improvement suggests that the LLM may necessitate additional training to effectively recognize out-of-distribution sounds. When we proceed to fully fine-tune both the audio encoder and the LLM (``Fully Fine-tuned"), the performance is enhanced by 0.09. This improvement suggests that the LLM is acquiring new audio concepts and refining the mapping from audio to text for novel classes.

Table \ref{incontext} compares the performance of different prompt strategies with the fully fine-tuned LTU model. Interestingly, supplying potential labels in the prompt to steer the model towards outputting labels of interest does not enhance the average performance. While some labels exhibit a slight improvement, others deteriorate, indicating that the model does not accord substantial attention to the information presented in the prompt.

Our final experiment with prompt modifications, which involved incorporating descriptions of Grouse calls into the prompt, also failed to meaningfully change the results. Although the AUC score for Grouse classification decreased by 0.06, the F-1 score rose by 0.05 to 0.46. Note, however, this score is still significantly lower than the performance achieved by the supervised model. The F-1 score was selected by calculating the F-1 for all threshold values with 0.001 increments on the validation set and choosing the threshold that yielded the optimal F-1 score for the reported test set results.

\subsection{Discussion}

We might expect that given the extensive training of the audio encoder, it would summarize the general properties of any audio input first as embeddings and then audio tokens. However, our results reveal a more complex interaction. The performance improvement observed when transitioning from partial to full fine-tuning suggests that the LLM is not merely processing the general audio properties encapsulated in the embeddings. Instead, it appears to focus on specific audio tokens that correspond to the related audio class, indicating that the LLM is mapping specific sound events to their associated words. This exposes a limitation. Ideally, the LLM should leverage the general properties in the embeddings to be able to reason with audio. However, its focus on specific audio tokens suggests it is not fully utilizing the information in the embeddings, potentially limiting its generalization capabilities. These findings underscore the challenges inherent in leveraging LLMs for in-context audio classification tasks, particularly when dealing with out-of-distribution sounds. They also highlight the potential benefits and limitations of various fine-tuning strategies and prompt modifications. While the LLM's ability to map specific sound events to associated words is beneficial, it is also a limitation in generalization to a wider range of audio inputs. The minimal impact of text prompts further suggests a lack of direct linkage between audio and text representations within the LLM.

\section{Experiment 2: Examining concept representations in an audio MLLM}
\label{exp2}

Building upon the findings of Experiment 1, our second experiment aims to delve deeper into the reasoning capabilities of MLLMs. Specifically, we are interested in discerning whether these models utilize information from the audio modality in their textual reasoning processes, or if they primarily map this information to individual keywords. We designed an input and expected output that necessitates the activation of reasoning abilities, such that we could modify the input to uncover what triggers these reasoning capabilities. 

While the interaction between concepts and their textual representations can manifest in myriad relations, one of the most structured and extensively studied is the semantic relation between words. This is meticulously catalogued in lexical databases such as WordNet \citep{miller1995wordnet}. WordNet organizes words into sets of synonyms called synsets, and records a variety of relations among these sets or their members, including synonyms, antonyms, hypernyms and hyponyms. This rich network of semantically related words and concepts provides a structured framework that can be used to understand and analyze the complex interrelationships between different concepts and their textual representations. Synonyms have previously been used to to evaluate vision MLLMs \citep[e.g.,][]{zohar2023lovm}, however the use of hypernyms is less common. Hypernyms have been used more widely in evaluating LLMs without a multimodal component \citep[e.g.,][]{shani2023towards} but are typically only mentioned in passing in evaluations of visual MLLMs \citep[e.g.,][]{chen2023can}.

We anticipated that when an LLM is augmented with additional modalities, such as audio, its capacity to answer questions pertaining to the relationships between concepts should extend to their corresponding representations within the new modality. This expectation is grounded in the understanding that the semantic relationships modeled by LLMs, such as hypernyms, are not confined to textual representations but can be extrapolated to other modalities as well.

\subsection{Methodology}
\label{exp2methodology}

\begin{table}
  \caption{Prompts used in experiment on concept representations in an audio MLLM for the two categories considered: similarity (synonyms) and hierarchy (hypernym). Slightly different prompts were crafted within each category for text- and audio-based queries.}
  \centering
  \begin{tabular}{lll}
    \toprule
    Similarity     & \emph{Text} & P1: Is \{concept\} similar to \{synonym\}?\\
    (synonym)     & \emph{Audio} & P2: Is the sound of the object in this audio signal similar to \{synonym\}?\\    
    \midrule
    Hierarchy     & \emph{Text} & P3: Is \{concept\} a type of \{hypernym\}?\\
    (hypernym)     & \emph{Audio} & P4: Is the sound of the object in this audio signal a type of \{hypernym\}?\\    
    \bottomrule
  \end{tabular}
  \label{exp2prompts}
\end{table}

LLMs model semantic relationships and can answer questions requiring reasoning, such as hypernyms and synonyms. Hypernyms represent a type of semantic relationship where one term serves as a broader category encompassing a set of other terms. For instance, `fruit' is a hypernym for `apple' and `orange'.
Synonyms represent another type of semantic relationship where two different terms share a similar meaning. We use this property of the hypernym and synonym relationships to construct the text prompt P3 from Table \ref{exp2prompts}. For synonyms, we pose the question ``Is {concept} similar to {synonym}?" which maps to the text prompt P1 from Table \ref{exp2prompts}. We expect LLMs to be able to answer these questions correctly if they understand the relationship between these concepts. This is because the ability to correctly identify and use hypernym and synonym relationships is a key aspect of understanding semantic relationships in language, and is a strong indicator of a system's ability to reason and generate meaningful responses in natural language processing tasks. The hypernym and synonym approaches can also be applied to the audio modality. If the model has learned the semantic relationship between the concepts from textual data, it should be able to transfer this understanding to audio data. Given the sound of a concept, the model should be able to identify it as a type of hypernym or as similar to a synonym, demonstrating its ability to reason.
To test this, we pose a question using an audio file that represents a concept. The question is framed as ``Is the sound of the object in this audio signal a type of {hypernym}?" for hypernyms, and ``Is the sound of the object in this audio signal similar to {synonym}?" for synonyms. These question formats correspond to the audio prompts P4 and P2 from Table \ref{exp2prompts}, respectively. For example, if we have an audio file of a songbird's chirping, the question would be ``Is the sound of the object in this audio signal a type of bird?", to which the answer should be yes. We also test a condition where a silent audio file is provided with the text prompt to assess whether the mere presence of audio changes the MLLM's reasoning processes.

We have constructed a concise benchmark that comprises 12 concept words. Each word is associated with up to 4 hypernyms, 4 synonyms, and 4 unrelated terms, culminating in a total of 111 word relationships (see Appendix \ref{SuppWords} for a full list).
We expect the model to respond favorably to all prompts involving concept hypernyms and synonyms, while showing a negative response to pairs of concepts and unrelated terms. For each word, we employ 4 audio files, each repeated 4 times, resulting in 16 queries. Consequently, we have 16 outputs per word pairs.To interpret these outputs, we employ regular expressions to discern whether the model's response aligns with the question, i.e., whether it is responding positively or negatively. A positive response, or an affirmation, is classified as a 'yes' response. We then compute the 'yes' rate, which is defined as the percentage of 'yes' responses out of the total 16 responses for each word pair. Notably, a 'no' response to an unrelated term is correct in our context and is thus flipped to 'yes' for consistency in the 'yes' rate, ensuring it accurately represents the model's correct identifications of both related and unrelated terms. Our samples are carefully selected from the evaluation set of AudioSet. We specifically opt for samples that contain only the target label, deliberately excluding any other sound events. A full list of the audio files we used is available in Appendix \ref{SuppAudio}. We use the same compute resources described for experiment 1, however we only run models in inference mode, adding up to less than 30 hours of GPU time.

\subsection{Results}

\begin{figure}
  \centering
  \includegraphics[width=0.9\textwidth]{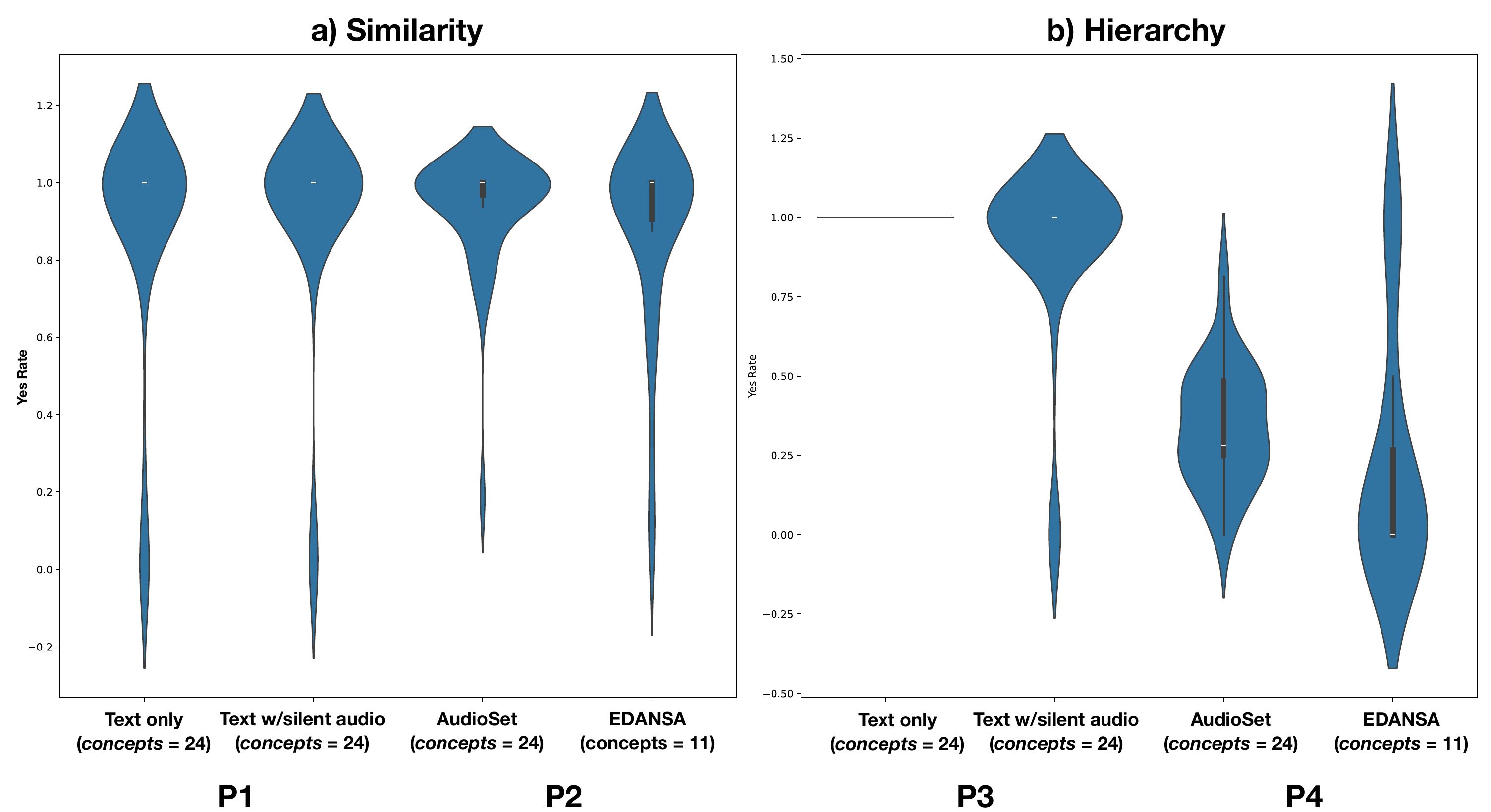}
  \caption{Results of experiment on concept representations in an audio MLLM. Subplot (a) shows the results for the similarity (synonyms and unrelated terms) category and subplot (b) for the hierarchy (hypernym) category. Each subplot shows violin plots of the yes rate for four different conditions: text-only prompting, text prompting with a silent audio file, text prompting with an audio file from AudioSet, and text prompting with an audio file from EDANSA. The prompts (P1–P4) are defined in in Table \ref{exp2prompts}.
  }
  \label{exp2results}
\end{figure}

Figure \ref{exp2results} presents the results of our experiments on concept representations in an audio MLLM. In the similarity category, the text-only approach and the text with silent audio approach both showed perfect performance. 
This indicates that the model was able to perfectly identify synonyms in a text-only context and when paired with silent audio. However, when actual audio data was introduced, there was a slight decrease in performance. The AudioSet condition had a median yes rate of 1.0 but with a slightly wider IQR of 0.031, indicating a small amount of variability in the responses. The EDANSA condition showed a further decrease in performance, with a median yes rate of 1.0 but an even wider IQR of 0.093, suggesting greater variability in the model's responses.

In the hierarchy category, the text-only approach and the text with silent audio approach again showed perfect performance.
This demonstrates the model's proficiency in identifying hypernyms in a text-only context and when paired with silent audio. However, the introduction of actual audio data resulted in a significant decrease in performance. The AudioSet condition had a median yes rate of 28.1\% and an IQR of 0.234, indicating a substantial amount of variability in the responses. The performance further deteriorated with the EDANSA condition, which had a median yes rate of 0.0 and an IQR of 0.265625.

\subsection{Discussion}

Our empirical findings show a distinct performance disparity in the Large Language Model (LLM) when tasked with answering questions pertaining to text-to-text versus audio-to-text relationships. In the similarity task, the model adeptly reasons about tokens, irrespective of their origin from audio captions or text, demonstrating its proficiency in identifying related words. However, a slight performance edge is observed in text-based tasks, indicating the model's stronger affinity for its native modality. In contrast, the hierarchy task presents a more complex challenge. While the LLM effectively leverages its reasoning capabilities with text, its performance wanes when presented with sound, suggesting a lack of connections between audio and textual concepts. This limitation is echoed in the findings from Experiment 1, where the model excelled in tasks involving one-to-one mapping, akin to the similarity task, but struggled with tasks requiring a broader understanding of relationships between concepts, as in the hierarchy task. We also observed that the model's performance is worse with EDANSA samples compared to AudioSet samples, indicating potential difficulties in processing and understanding out-of-distribution sound data.

\section{Limitations}
\label{limitations}

The experiments in this paper are exploratory and limited in scope. Most notably, the Grouse-specific components of Experiment 1 used only the prompts described in Appendix \ref{SuppPrompt}. Although this is a carefully crafted prompt, there remains the possibility that either the wording or content of the prompt was suboptimal. Similarly, Experiment 2 is limited to the carefully curated words and audio files in Appendices \ref{SuppWords} and \ref{SuppAudio}, however, the amount of between-group consistency shown in the plots in Figure \ref{exp2results} for the synonymy experiment as well as the significant difference between groups for the hypernym experiment suggest that the dataset was sufficiently sized to capture the behavior related to LLM context representations that we were interested in exploring. A final limitation is our use of a single audio MLLM, LTU. This was largely constrained by available compute resources, but given the consistency in architecture across the three most currently popular audio MLLMs (Pengi, LTU, and SALMNON), it is likely that we would find similar results using any of them. 

\section{Conclusions and Future Work}
\label{conclusions}
In this paper, we evaluated whether an audio MLLM, specifically LTU, can exploit the reasoning capabilities of LLMs to learn in context through prompting (Experiment 1). We demonstrated the limitations of a current audio MLLM in leveraging its LLM's reasoning power via in-context prompting. This led to the design and implementation of Experiment 2 to examine the audio MLLM's concept representations with synonyms and hypernyms, demonstrating that the audio MLLM is not fully integrating text and audio information in a way that it can perform hierarchical-related reasoning on audio input. A common solution to address this reasoning in vision MLLMs is generating additional data pairs of text and images that require the model to pay attention to the missing ability, such as the order of the items in the images \citep{yuksekgonul2022}. This is a limited solution, which would only solve reasoning on tasks covered by the data pairs. There is potential to improve reasoning through finer-grained alignment between the modalities, although this is still an open area of research.

A better understanding of LLMs' reasoning capabilities, both in isolation and in the context of MLLMs, has the potential for broader societal impacts. This includes contributing to decoding what LLMs are and are not modeling, which can help delineate some of the known limitations that should be considered in their use. This could be used either positively or negatively, but we anticipate the development of such techniques would largely have a positive impact. This work is also the first step towards understanding how to leverage domain-specific description-based knowledge in audio classification with MLLMs, as exemplified by our in-context prompting experiment on grouse calls, to better proxy how humans learn.

\begin{ack}
This material is based upon work supported by the National Science Foundation under Grants No. 1839185 and 2228910.

\end{ack}

\bibliographystyle{plainnat}
\bibliography{refs}

\medskip

\newpage

\appendix

\section{Prompt used for In-Context Learning sub-experiment in Experiment 1}
\label{SuppPrompt}

\begin{verbatim}
# source for Grouse description:
# https://birdsoftheworld.org/bow/species/wilpta/cur
# /introduction#vocal
{
    'default':
        'write an audio caption describing the sound',
    'label_in_prompt':
        'Write an audio caption describing the sound. \
   Could the sound be Silence, Biological soundscape,\
   aviation noise, Rainfall, Grouse, Insect,\
   Songbird, Duck and/or Goose and/or Swan,\
   anthropogenic noise, physical soundscape, Bird, or Wind?',
    'grouse01':
        """provide labels for the audio file, could it be a Grouse?
here is detailed information on Grouse call types: 
Both Sexes. (1) Kok is a short (50 ms) clucking call; \
volume and frequency vary with intensity of arousal\
(2) Ko-ko-ko is a low-amplitude call given \
in prolonged bouts, sounding like low growls; \
(3) Krrow is a medium-length (50-300 ms)\
call that rises quickly and falls slowly in frequency: In males, \
sounds like bugow; in females, like meow; given during aggressive \
disputes \
(4) Kohwa, Kohway, and Kohwayo are medium-length (100 ms) \
calls often given in association with \
Krrow during aggressive interactions. \
(5) Aroo is a variable call with falling \
and rising pattern of frequency modulation at varying rate and \
intensity. \
(6) Rattle (song on the ground) is a long (800 ms), accelerating \
string of short elements similar to Kok\
(7) Flight Song (Aerial Bek) is a 2-part vocalization; first part \
is a decelerating series of modified Ko-ko-ko calls, second part \
a decelerating Kohwa  alternatively described as a series of nasal \
barks, typically a few single notes followed by a rattling sequence \
of 8–12 guttural notes, and several doubled notes on landing,
“whek!..whek-kekekrrrrekek-kek...koh-wa..koh-wa..koh-wa". \
(8) Scream is a brief, high-frequency\
call with indeterminate harmonic structure. (9) Hiss is a band of \
white noise about 2 s long.\
list labels for the audio file:
""",
    'grouse02_label_in_prompt':
        """provide labels for the audio file, could it be a Grouse?
here is detailed information on Grouse call types: 
Both Sexes. (1) Kok is a short (50 ms) clucking call; \
volume and frequency vary with intensity of arousal\
(2) Ko-ko-ko is a low-amplitude call given \
in prolonged bouts, sounding like low growls; \
(3) Krrow is a medium-length (50-300 ms)\
call that rises quickly and falls slowly in frequency: In males, \
sounds like bugow; in females, like meow; given during aggressive \
disputes \
(4) Kohwa, Kohway, and Kohwayo are medium-length (100 ms) \
calls often given in association with \
Krrow during aggressive interactions. \
(5) Aroo is a variable call with falling \
and rising pattern of frequency modulation at varying rate and \
intensity.
(6)  Rattle (song on the ground) is a long (800 ms), accelerating \
string of short elements similar to Kok\
(7) Flight Song (Aerial Bek) is a 2-part vocalization; first part \
is a decelerating series of modified Ko-ko-ko calls, second part \
a decelerating Kohwa  alternatively described as a series of \
nasal barks, typically a few single notes followed by a rattling  \
sequenceof 8–12 guttural notes, and several doubled notes on \
landing, “whek!..whek-kekekrrrrekek-kek..koh-wa..koh-wa..koh-wa".\
(8) Scream is a brief, high-frequency\ call with indeterminate \
harmonic structure.\
(9) Hiss is a band of white noise about 2 s long.\
 Write an audio caption describing the sound. \
 Could the sound be Silence, Biological soundscape,\
 aviation noise, Rainfall, Grouse, Insect,\
 Songbird, Duck and/or Goose and/or Swan,\
 anthropogenic noise, physical soundscape, Bird, or Wind?'
"""
}

\end{verbatim}

\section{Alternative labels considered in Experiment 1}
\label{alternative_labels}

\begin{table}[h]
\centering
\begin{tabular}{p{3cm}p{10cm}}
\toprule
\textbf{Original label} & \textbf{Alternatives} \\ 
\midrule
Silence & Silence \\
Biophony & Biological soundscape, animal chorus, wildlife sounds, ecosystem acoustics \\
Aircraft & airplane, aviation noise, air traffic noise \\
Rain & Rainfall, raindrops, rain pattering \\
Grouse & Grouse \\
Bug & Insect, bug, entomological sounds, insect calls \\
Songbird & Songbird \\
DGS & Duck and/or Goose and/or Swan \\
Anthropophony & human-made noise, industrial noise, anthropogenic noise \\
Geophony & natural ambient sounds, non-biological soundscape, physical soundscape \\
Bird & Bird \\
Wind & Gust sounds, blowing wind, aeolian sound \\
\bottomrule
\end{tabular}
\label{tab:alternative_mapping}
\end{table}

\section{Words used for concepts in Experiment 2}
\label{SuppWords}

\begin{longtable}{p{2.5cm} p{2.5cm} p{3cm} p{3cm} p{3cm}}
\hline
\textbf{Category} & \textbf{Label} & \textbf{Synonym(s)} & \textbf{Hypernym(s)} & \textbf{Unrelated} \\
\hline
\endfirsthead

\hline
\textbf{Category} & \textbf{Label} & \textbf{Synonym(s)} & \textbf{Hypernym(s)} & \textbf{Unrelated} \\
\hline
\endhead

\hline
\endfoot

\hline
\endlastfoot

biophony & bird & fowl & vertebrate & speech \\
         &      & avian & craniate & speaking \\
         &      & aves & chordate & wind \\
         &      &  & animal & breathing \\
\hline
biophony & cattle & cows & bovine & working \\
         &        & oxen & bovid & grumbly \\
         &        & bos taurus & ruminant & melodic \\
         &        &  & animal & wind \\
\hline
biophony & dog & canis familiaris & canine & comforting \\
         &     & domestic dog & canid & speech \\
         &     &  & domestic animal & brief \\
         &     &  & animal & music \\
\hline
biophony & insect & bug & arthropod & wind \\
         &        &  & invertebrate & authoritative \\
         &        &  & animal & characterized \\
         &        &  &  & speech \\
\hline
anthrophony & aircraft & airplane & craft & speech \\
            &          & airship & vehicle & natural \\
            &          & aeroplane & conveyance & male \\
            &          &  & transport & rich \\
\hline
anthrophony & car & motorcar & motor vehicle & speech \\
            &     & automobile & vehicle & police \\
            &     & auto & conveyance & generic \\
            &     & machine & transport & music \\
\hline
anthrophony & fireworks & pyrotechnics & low explosive & speech \\
            &           &  & explosive & speaking \\
            &           &  &  & warm \\
            &           &  &  & generic \\
\hline
anthrophony & alarm & alert & signal & speaking \\
            &       &  & sign & car \\
            &       &  &  & vehicle \\
            &       &  &  & authoritative \\
\hline
geophony & rain & rainfall & precipitation & surface \\
         &      & rainwater & downfall & thunder \\
         &      &  & weather & human \\
         &      &  & atmospheric condition & authoritative \\
\hline
geophony & wind & air current & weather & microphone \\
         &      & current of air & weather condition & male \\
         &      &  & atmospheric condition & rich \\
         &      &  &  & instrument \\
\hline
geophony & thunder & boom & thunderstorm & rolling \\
         &         &  & electrical storm & speech \\
         &         &  & storm & footsteps \\
         &         &  & atmospheric phenomenon & whistling \\
\hline
geophony & waterfall & falls & water & speech \\
         &            &  &  & male \\
         &            &  &  & music \\
         &            &  &  & man \\
\hline

\end{longtable}

\section{Concepts and audio files used in Experiment 2}
\label{SuppAudio}

\begin{longtable}{p{1.5cm} p{1.5cm} p{2.5cm} p{9cm}}
\hline
\textbf{Category} & \textbf{Label} & \textbf{AudioSet ID} & \textbf{EDANSA\_IDs} \\
\hline
\endfirsthead

\hline
\textbf{Category} & \textbf{Label} & \textbf{AudioSet ID} & \textbf{EDANSA\_IDs} \\
\hline
\endhead

\hline
\endfoot

\hline
\endlastfoot

biophony & bird & -XilaFMUwng & INP-AR-03\_20190617\_220000\_8m\_30s\_\_8m\_40s \\
         &      & -qS77R0Y1K8 & S4A10227\_20190611\_043000\_22m\_19s\_\_30m\_34s\_splt-21 \\
         &      & 12T-9dLEbY8 & S4A10301\_20190613\_000000\_12m\_50s\_\_13m\_0s \\
         &      & 1dH-lZ8TNLU & S4A10301\_20190613\_000000\_7m\_30s\_\_7m\_40s \\
\hline
biophony & cattle & sbpW3Z87Nbc &  \\
         &        & z3YihIejSIA &  \\
         &        & UYBuKiXo92s &  \\
         &        & KksMNKXuiNw &  \\
\hline
biophony & dog & 20qZLse0acs &  \\
         &     & 8CrTpWNBiTo &  \\
         &     & E6QQRZHrx6s &  \\
         &     & KRdvyjpQfoI &  \\
\hline
biophony & insect & 9j\_FItO0jt8 & SINP03/SINP-03\_20190704\_210000\_1m\_30s\_\_1m\_40s \\
         &        & zPSH6-UC4Og & S4A10327\_20190725\_104602\_45m\_50s\_\_46m\_0s \\
         &        & 5j\_v9dhjbdU & anwr\_41\_S4A10273\_20190707\_183000\_exact\_2019-07-07 \\
         &        & QBj5dyzsJkY & SINP03/SINP-03\_20190708\_173000\_15m\_20s\_\_15m\_30s \\
\hline
anthrophony & aircraft & -OVb-UG8yJw & S4A10361\_20210515\_010002\_41m\_30s\_\_41m\_40s \\
            &          & -ocADGlyaHc & S4A10272\_20190509\_073000\_39m\_20s\_\_39m\_30s \\
            &          & 7S88FsFE5EE & 18/2019/S4A10280\_20190525\_104602\_33m\_22s\_\_33m\_32s \\
            &          & DU3cNZdlylQ & S4A10298\_20210730\_060002\_55m\_50s\_\_56m\_0s \\
\hline
anthrophony & car & xRonpWC3SvY & S4A10443\_20200428\_100412\_2m\_0s\_\_2m\_10s \\
            &     & -aOxR6ILsw8 & S4A10291\_20191010\_144000\_2m\_0s\_\_2m\_10s \\
            &     & 4TshFWSsrn8 &  \\
            &     & Kwpn3utYEHM &  \\
\hline
anthrophony & fireworks & L6QtigLJD\_4 &  \\
            &           & l7RTgupQWcc &  \\
            &           & UxEyOSK9nxo &  \\
            &           & AJRD-zU2Akw &  \\
\hline
anthrophony & alarm & 3o-q-VMhyA8 &  \\
            &       & FBut7W5XwnA &  \\
            &       & T\_FZMsRHzLc &  \\
            &       & fcsGkE89Qi8 &  \\
\hline
geophony & rain & 96HJ2f5dj6U & S4A10273\_20190803\_050000\_42m\_24s\_\_57m\_34s\_splt-29 \\
         &      & fvQeqBqqcVw & S4A10273\_20190803\_050000\_42m\_24s\_\_57m\_34s\_splt-53 \\
         &      & johz0yXuORc & AR01/2018/INP-AR-01\_20180817\_020000\_7m\_51s\_\_8m\_1s \\
         &      & fwas0HLGbqM & S4A10287\_20190803\_050000\_rain02\_splt-2 \\
\hline
geophony & wind & A74lbeD1k1o & S4A10273\_20190803\_093000\_55m\_0s\_\_56m\_50s\_splt-2 \\
         &      & CkutJYIfghs & anwr\_37\_S4A10279\_20190603\_043000\_exact\_2019-06-03\_04-38-36\_0m\_0s\_\_0m\_10s \\
         &      & AkUDv7JexjQ & S4A10295\_20190708\_000000\_49m\_50s\_\_50m\_0s \\
         &      & zzbTaK7CXJY & dempster/25/2020/S4A10334\_20200415\_140002\_2m\_54s\_\_3m\_4s \\
\hline
geophony & thunder & 0439dMJj-FY &  \\
         &         & przrSPZgOkY &  \\
         &         & ZBaYrfz5afo &  \\
         &         & 54wNjdYr8ww &  \\
\hline
geophony & waterfall & FF2bhR7s3VY &  \\
         &            & JfDeETDDwhM &  \\
         &            & VMbJTgzMhKE &  \\
         &            & hfIfBPkH8Fo &  \\
\hline

\end{longtable}

\end{document}